\documentstyle[aps,multicol,epsfig,eqsecnum]{revtex}

\newcommand{\bR}{\textbf{R}}
\newcommand{\dq}[1]{{\!{\text d}^2}#1}
\newcommand{\ux}{{\underbar{x}}}
\newcommand{\intds}{\int\limits_0^N \!{\text d} s\,}
\newcommand{\intdsp}{\int\limits_0^N \!{\text d} {s^\prime}\,}
\newcommand{\drds}{\left({\partial\bR(s)\over\partial s}\right)^2}

\begin{document}
\large
\title{ Polymer adsorption on heterogeneous surfaces}
\author{Gregor Huber,  Thomas A.Vilgis}
\address{Max-Planck-Institut f\"ur Polymerforschung, Postfach 3148,\\
D-55021 Mainz, Germany}
\date{\today}
\maketitle

\begin{abstract}
  The adsorption of a single ideal polymer chain on energetically
  heterogeneous and rough surfaces is investigated using a variational
  procedure introduced by Garel and Orland ({\sl Phys.~Rev.~B} {\bf 55}
  (1997), 226).  The mean polymer size is calculated perpendicular and
  parallel to the surface and is compared to the Gaussian conformation and to
  the results for polymers at flat and energetically homogeneous surfaces. The
  disorder-induced enhancement of adsorption is confirmed and is shown to be
  much more significant for a heterogeneous interaction strength than for
  spatial roughness.  This difference also applies to the localization
  transition, where the polymer size becomes independent of the chain length.
  The localization criterion can be quantified, depending on an effective
  interaction strength and the length of the polymer chain.
\end{abstract}


\section{Introduction}

\begin{figure}
\begin{center}
\begin{minipage}{12cm}
\centerline{{\epsfig{file=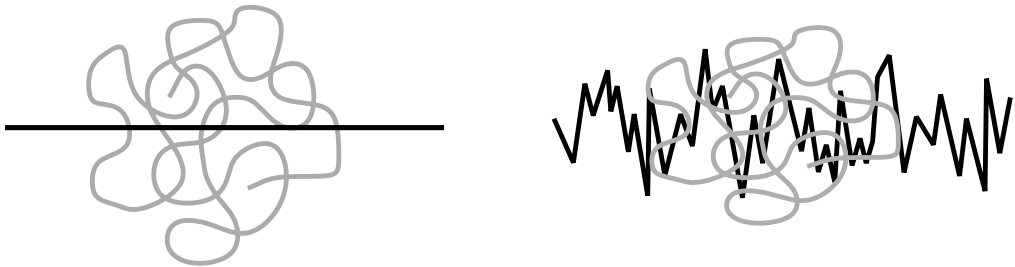,height=3.5cm}}} 
\vspace*{10pt}
\caption{\label{fig:flory1} Sketch of the main mechanism of adsorption
    enhancement by surface roughness: the number of possible
    binding sites increases
    without being balanced by a loss in configurational entropy.}
\end{minipage}
\end{center}
\end{figure}

The adsorption of polymers on flat and homogeneous attractive surfaces has
been the subject of many investigations, see e.g.\ \cite{Eisen,Fleer}, but
naturally occurring surfaces usually are rough and/or energetically
inhomogeneous, the heterogeneity leading to an enhancement of adsorption under
quite general conditions
\cite{HV95,VH94,Linden,Andelman,Baum2,BBB,BBB2,Douglas,HJP}: already simple
physical arguments contain the statement that upon increasing the surface
irregularity, the number of polymer-surface interactions is strongly enhanced
relative to the idealized planar surface (see figure \ref{fig:flory1}).  This
is a consequence of a larger probability of polymer-surface intersection with
increasing roughness.

The study presented here is partly motivated by the theoretical investigation
of reinforcement mechanisms in carbon black filled elastomers, where the
polymer adsorption is substantial as there is a strong binding of the polymers
to the surface (leading to a layer of localized polymers, the so called "bound
rubber" \cite{DBW,Mei}).  These surfaces are rough and even fractal on many
decades of their size, down to the molecular size range, as well as highly
energetically (i.e.~chemically) heterogeneous: the distribution of interaction
strengths can be characterized by high energy sites surrounding a relatively
low energy background.  Therefore in principle both kinds of disorder should
be incorporated in a theory which is supposed to explain the bound rubber
phenomenon.
 
In the literature so far spatial and energetical heterogeneities were always
treated separately.  In this paper we want to present a treatment which covers
both sorts of surface heterogeneity, thereby making possible a comparison
concerning the strength of adsorption enhancement.  Thus the general aim of
the present paper is to produce a theory for the adsorption and localization
of polymers on such surfaces. This is a nontrivial problem.  For flat surfaces
the description of ideal polymers by Schr\"odinger type equations is regarded
as the method of choice, but for rough surfaces this method cannot be used in
general, because the equations can be solved only for some highly specific
boundary conditions, e.g., regular fractal surfaces \cite{BBB}.  Here we want
to investigate the polymer behavior near surfaces with any type of surface
disorder, as well as surfaces with a wide range of heterogeneities in the
attractive potential.  Therefore a path integral formalism seems to be
appropriate, since the surface disorder can be dealt with explicitly up to an
advanced stage of the calculation and disorder averages are less difficult.

However, some technical problems appear, forcing us to restrict the
investigation to an idealized model system: a single ideal polymer chain in
interaction with an attractive, penetrable, infinite surface.  
The treatment of real chains would require field theoretical approaches
\cite{Eisen,Diehl} which are beyond the scope of the present paper.  The
penetrability of the surface (both sides of the surface are accessible for the
polymer), which is of interest in the context of polymer chains at membranes
or interfaces, here has to be assumed in order to avoid complex boundary
conditions.  In the problem of polymer adsorption on flat surfaces,
non-penetrability is relatively easy to handle: the polymers are considered in
half space \cite{Eisen,Fleer}. For random surfaces an half space treatment is
no longer possible. Therefore we will loose information on the localization
behavior, but nevertheless expect general insight into the problem.

The results of the calculations are expressions for the size of the chain,
depending on the mean interaction strength $w_0$, the chain length $N$ and the
disorder parameters.  The paper is organized as follows.  The size of the
chain $R_{\bot}$ perpendicular to the surface is obtained
\begin{itemize}
\item[(i)] in section \ref{scaling} for strong spatial disorder, using a
  scaling argument \`{a} la de Gennes \cite{dG} -- this calculation intends to
  give a better insight into the main mechanism of adsorption enhancement;
\item[(ii)] in section \ref{vario} for weak spatial disorder and energetically
  heterogeneous surfaces, using a variational calculation.
\end{itemize}
In the latter case, first the Hamiltonian is formulated which explicitly
contains the surface geometry and a distribution of interaction strengths.
Then the free energy is approximated using a Feynman variational procedure.
It turns out that for our problem an extension of a trial Hamiltonian
introduced by Garel and Orland \cite{Garel} is much more suitable than the
replica trial Hamiltonian frequently used in connection with polymers in
disordered environments \cite{EM,EC,CB}.  In comparison with Garel and\ 
Orland, our main modification is the separation of the monomer coupling
parameter into components of different space direction, thus enabling the
polymer size parallel and perpendicular to the surface normal to be
distinguished.  After minimization of the free energy, the effect of surface
heterogeneity can be summarized in an effective interaction strength which in
most cases is larger than the mean interaction strength.  The variational
scheme is first tested on a flat and homogeneous surface, where it is shown
that the result agrees with the scaling estimate and the results given
earlier \cite{Eisen,Fleer} for ideal chains.  Then the effective interaction
strength is calculated for the special cases of a periodic and random
distribution of interaction strength on a flat surface and for a periodic and
random surface profile at constant interaction strength, respectively.  This
also allows us to consider the localization transition.
The results are discussed and summarized in section \ref{conc}.

\vfill\break


\section{Scaling Argument for Fractal Spatial Disorder}\label{scaling}

\subsection{Flat Surface}

First we briefly review a simple scaling treatment of an ideal chain adsorbed
on a flat surface, as introduced by de Gennes \cite{dG}.
Let $R_\bot$ and $R_{||}\simeq R_0\simeq bN^{1/2}$ be the mean size of an
ideal polymer (with $N$ monomers and effective monomer length $b$)
perpendicular and parallel to the surface, respectively.  The monomer density
is assumed to be constant in a region of size $R_\bot R_{||}^2$.  Then
the number $\cal N$ of monomers bounded to the surface is estimated as 
\begin{equation}
{\cal N}\ =\ b R_{||}^2\; {N\over R_\bot R_{||}^2}\ =\ {b\,N\over R_\bot}.
\end{equation}
Consequently the free energy can be written as
\begin{equation}
  \beta F\ \approx\ {R_0^2\over R_\bot^2} - \beta w\,{\cal N}\ =\ 
  {b^2N\over R_\bot^2} - \beta w\;{b N\over R_\bot},
\end{equation}
the first term being the confinement energy, the second one due to contact
interactions with the surface ($-w$ is the effective monomer attraction,
$\beta$ the inverse temperature).  Minimization of the free energy ${\partial
  F/\partial R_{\bot}}=0$ gives an expression for the polymer thickness
$R_{\bot}$ perpendicular to the surface,
\begin{equation}\label{flat}
  R_{\bot} \simeq {b \over \beta w}.
\end{equation}
Thus the thickness of the polymer reduces with growing attractive interaction
strength, as expected.  The independence of the chain length $N$ indicates
that the polymer is in the so called ''localized'' regime \cite{CB}.

\subsection{Generalization for Fractal Surfaces}

A fractal surface may be characterized by its fractal dimension $d_S$, $2\leq
d_S\leq 3$, a value $d_S=2$ corresponding to a flat surface. The limit $d_S\to
3$ produces an extremely rough, space-filling surface, Brownian surfaces
\cite{Feder} are characterized by $d_S=2.5$.

Now the number of bounded monomers is written as
\begin{equation}
  {\cal N} =\ b^{3-d_S}\,R_{\bot}^{d_S}\;{N\over R_{\bot}^3}.
\end{equation}
Running through the same procedure as above yields
\begin{equation}
  R_{\bot}\simeq{b\over (\beta w)^{1/(d_S-1)}},
\end{equation}
so that the result (\ref{flat}) for the case of a flat surface $d_S=2$ is
recovered.

From (\ref{flat}) we have $\beta w<1$ because of $b\ll R_{\bot}$ for weak
adsorption, where no complete collapse on the surface takes place. In fact for
most materials values about $\beta w\sim 0.01\dots 0.1$ are found
\cite{Klein}.  Therefore the polymer adsorption on rough surfaces ($d_S>2$)
generally is enhanced compared to the case of a flat surface, i.e.
$R_{\bot}^{\text{rough}}<R_{\bot}^{\text{flat}}$.

Although this is a crude argument, it gives an insight into the main aspects
of adsorption enhancement: the crucial point is the competition between the
gain in potential energy obtained by binding to the surface and the loss in
chain entropy associated with the confinement of chains in comparison to free
chains.  Therefore,  the dominating mechanism in our consideration above is the
increasing number of binding sites at a rough surface without paying an
entropy penalty, which means that a chain has to lose less configurational
entropy to adsorb on a rough surface.
This is in agreement with results of much more expensive previous calculations
by Douglas et al.\ \cite{Douglas} and Hone et al.\ \cite{HJP}.

A similar argument holds for the case of energetical heterogeneity \cite{BM}:
with a distribution of the interaction strength on the surface, the chain can
select the strong binding points without changing its configuration too much,
thus seeing a larger effective interaction strength.

\section{Variational Calculation} \label{vario}

For a systematic study of $R_{\bot}$ in the case of spatial and energetical
heterogeneity, the free energy is calculated via a variational procedure,
where the disorder is treated as a quenched (i.e.\ frozen) randomness. The
replica method is avoided by introduction of an additional variational
parameter, see next section.  We consider an ideal chain at an infinite,
penetrable, well defined surface with low profile.  Furthermore we assume an
attractive contact (i.e.\ extremely short range) interaction between chain and
surface that can be mimicked by a delta-potential.

Now this system is represented by its Edwards-Hamiltonian, which reads
\begin{equation} 
\beta H = {3\over 2b^2} \intds \drds  + \beta \intds V(\bR(s)),
\end{equation}
$\bR(s)$ being the chain segment vector position.  The potential contains the
polymer-surface coupling,
\begin{equation} 
V(\bR(s)) = -\int\dq x\, K[h(\ux)]\; b\,w(\ux)\, 
            \delta(\bR(s) - {\textbf{h}}(\ux))
\end{equation}
with ${\textbf{h}}(\ux)=(\ux,h(\ux))$, where $\ux=(x_1,x_2)$ is an internal
surface vector. The surface disorder is described by $w(\ux)$ for energetical
disorder, i.e.\ an interaction strength varying on the surface, and by
$h(\ux)$ for spatial disorder, i.e.\ a rough surface profile.  The factor
$K[h(\ux)]=(1+|\nabla h(\ux)|^2)^{1/2}$ takes account of the local deflection
of the surface in cartesian coordinates.

In order to approximate the free energy we make use of a Feynman variational
procedure,
\begin{equation}
\left\langle e^{-\beta (H-H_0)} \right\rangle_{\!H_0} 
\geq e^{-\beta\left\langle H-H_0\right\rangle_{\!H_0}},
\end{equation}
where $\left\langle...\right\rangle_{\!H_0}$ denotes the average with respect
to a trial Hamiltonian $H_0$.  This gives an upper bound to the free energy 
\begin{equation} \label{betaF} 
\beta F\leq \beta F^{*} 
= \beta F_0 + \beta \left\langle H-H_0 \right\rangle_{\!H_0}
\end{equation}
with the abbreviation
\begin{equation}
  \beta F_0 = \log\left(\int{\cal D}\bR \exp\{-\beta H_0\}\right).
\end{equation}
$\beta F^{*}$ has to be minimized to give the best estimate for the
true free energy $\beta F$.

\subsection{Trial Hamiltonian}

The appropriate choice of the trial Hamiltonian is most significant for the
utility of the variational procedure.  Here we take an extension of a form
suggested by Garel and Orland \cite{Garel},
\begin{equation} 
\beta H_0 = {1\over 2} \sum_{j=1}^3 \intds \intdsp 
(R_j(s)-B_j) g_j^{-1}(|s-s^\prime|) (R_j(s^\prime)-B_j).
\end{equation}

Its features are: (a) it is quadratic in $\bR(s)$, so that an exact
calculation of $\beta F_0$ is possible; (b) the coupling of chain segments is
mediated by the variational parameters $g_j(|s-s^\prime|)$, one for each
direction of space: the indices 1 and 2 are identified with the coordinates
$x_{1}$ and $x_{2}$ of the surface parameterization, index 3 corresponds to
the $z$ coordinate parallel to the average surface normal; (c) there is an
additional variational parameter $\textbf{B}$, equivalent to a translation of
the centre of mass of the chain. It should be mentioned that this type
of variational principle was originally designed to avoid replica theory in
random systems \cite{Garel}. This is another reason why the Garel-Orland
method is chosen here. If the polymer is assumed to stick permanently at some
place along the disordered surface, the problem falls into classes dealing
with "quenched disorder" and difficulties with replica symmetry breaking
arise. In the following we will show that the Garel-Orland method is indeed
useful to treat the problem of polymer adsorption on disordered surfaces as it
yields physically sensible results.

Assuming cyclic boundary conditions $\bR(N)\equiv\bR(0)$, the variational free
energy (\ref{betaF}) can now be calculated to give
\begin{equation}
\beta F^{*} = - \sum_{n=1}^\infty \sum_{j=1}^3\log {\tilde{g}_j(n)\over b^2}
       + \sum_{n=1}^\infty \sum_{j=1}^3 N\omega_n^2 {\tilde{g}_j(n)\over b^2} 
              + \beta {\cal W}(\textbf{B},\textbf{G})
\end{equation}
with $\omega_n = 2n\pi/N$. Here the interaction energy ${\cal
  W}(\textbf{B},\textbf{G})$ is the only term depending on the interaction
potential,
\begin{eqnarray}
{\cal W}(\textbf{B},\textbf{G}) 
&=& {-Nb\over(2\pi)^{3/2}(G_1G_2G_3)^{1/2}}\, \int\dq x\ 
     K[h(\ux)] \, w(\ux)\, \exp\left\{ - \sum_{i=1}^3
     {(B_i-h_i(\ux))^2  \over 2G_i}\right\},
\end{eqnarray}
the $h_i(\ux)$ being the components of ${\textbf{h}}(\ux)$, i.e.\ 
$h_3(\ux)\equiv h(\ux)$.  The parameters $G_j$ are defined by
\begin{equation}
G_j = 2 \sum_{n=1}^\infty \tilde{g}_j(n) 
       = {2\over N}\sum_{n=1}^\infty\intds\cos(\omega_n s) g_j(s),
         \qquad j=1,2,3. 
\end{equation}
$G_j$ can be identified with the mean square radius of the polymer parallel
($G_3$) or perpendicular ($G_{1}$, and $G_{2}$) to the surface normal.

\subsection{Minimization of the Free Energy}

Following the lines of Garel and Orland, the minimization of $\beta F^{*}$
with respect to $\tilde{g}_j(n)$ and $\textbf{B}$ leads to
\begin{equation} \label{MinB}
{\bf\nabla}_\textbf{B}{\cal W}(\textbf{B},\textbf{G}) \stackrel{!}{=} {\bf 0}
\end{equation}
and
\begin{equation} \label{MinG}
\tilde{g}_j(n) \stackrel{!}{=} {b^2 \over N\omega_n^2 + \beta b^2 
\displaystyle{\partial^2 {\cal W}(\textbf{B},\textbf{G}) \over\partial B_j^2}}
\end{equation}
because of $\partial{\cal W}(\textbf{B},\textbf{G})/(\partial \tilde{g}_j(n))
= \partial^2{\cal W}(\textbf{B},\textbf{G})/(\partial B_j^2)$.  As already
discussed by Garel and Orland \cite{Garel}, one does in general expect the
variational equations to have several solutions. This especially applies to
(\ref{MinB}), since we consider an infinite surface, e.g.\ leading to an
infinite number of solutions in the case of a periodic surface heterogeneity.
All these solutions have equal free energy.

Introducing the abbreviation
\begin{equation} 
\alpha_j = \left( {N^2\, b^3 \over 4(2\pi)^{1/2}}\, 
                  {\beta |w_j^{\text{eff}}| \over G_j^{3/2}}
         \right)^{1/2},
\end{equation}
the optimized parameter $G_j$ is calculated from (\ref{MinG}) as
\begin{equation} \label{Gallgemein}
G_j = 
\left\{\begin{array}{l@{\quad{\mathrm{for}\ \ }}l}
      \displaystyle{Nb^2\over4}\ 
      \displaystyle{\coth(\alpha_j)-\alpha_j^{-1}\over \alpha_j}
      &w_j^{\text{eff}}\geq 0,\\[10pt]
       -\displaystyle{Nb^2\over4}\ 
      \displaystyle{\cot(\alpha_j)-\alpha_j^{-1}\over \alpha_j}
      &w_j^{\text{eff}}< 0.
      \end{array}\right. 
\end{equation}
The effective interaction strength $w_j^{\text{eff}}$ contains all relevant
surface and polymer properties and is given by
\begin{equation} \label{wjdef}
w_j^{\text{eff}} =  {(2\pi)^{1/2} G_j^{3/2} \over N\, b}\ 
            \left. {\partial^2 {\cal W}(\textbf{B},\textbf{G}) \over
                   \partial B_j^2} \right|_{{\bf\nabla}_{\bf B}{\cal W}
                                         (\bf{B},\bf{G}) = {\bf 0}}
\end{equation}
In two special cases results can be obtained very easily: 
\begin{itemize}
\item[(i)] if there is no interacting surface at all, i.e.\ $w(x)\equiv 0$,
  then we immediately have $\alpha_j=0$ and therefore $G_j = Nb^{2}/12 \equiv
  R_0^2/2$, the chain conformation is purely Gaussian in all directions, as
  expected;
\item[(ii)] for an ideal surface, which means $w(x)\equiv w_0$ and
  $h(\ux)\equiv h_0$, the effective interactions strengths are calculated as
  $w_{1/2}^{\text{eff}} = 0$ and $w_3^{\text{eff}} = w_0$.  So the definition
  (\ref{wjdef}) of $w_j^{\text{eff}}$ guarantees correct results for this
  case.
\end{itemize}
The explicit form of $w_j^{\text{eff}}$ for various special sorts
of surface heterogeneity is calculated in the next section.

The discussion of (\ref{Gallgemein}) is complicated by the fact that the
effective interaction strength itself is a function of the polymer extensions
in different directions.  But in general (\ref{Gallgemein}) can for
$w_j^{\text{eff}}\geq 0$ be expanded in the limits of small and large
$\alpha_j$. This yields the mean polymer extension into the different
directions of space, $\bar{R}_3$ being parallel to the surface normal,
$\bar{R}_1$ and $\bar{R}_2$ perpendicular to it, if $\langle h(\ux)\rangle =0$
is assumed,

\begin{equation} \label{Rj}
  \bar{R}_j \simeq \sqrt{G_j} \simeq 
  \left\{\begin{array}{l@{\quad{\mathrm{for}\ \ }}l}
   R_0 \left\{1-c N^{1/2} \beta w_j^{\mathrm{eff}}\right\} 
     & \beta w_j^{\text{eff}} \ll N^{-1/2} \\[10pt]
   \displaystyle{b\over \beta w_j^{\mathrm{eff}}} 
   \left\{ 1 - \displaystyle{\pi \over N (\beta w_j^{\text{eff}})^2}\right\}
     & \beta w_j^{\text{eff}} \gg N^{-1/2}
         \end{array}\right. 
\end{equation}
where, as above, $R_0$ denotes the radius of gyration of the corresponding
Gaussian chain.

\begin{figure}
\begin{center}
\begin{minipage}{12cm}
  \centerline{{\epsfig{file=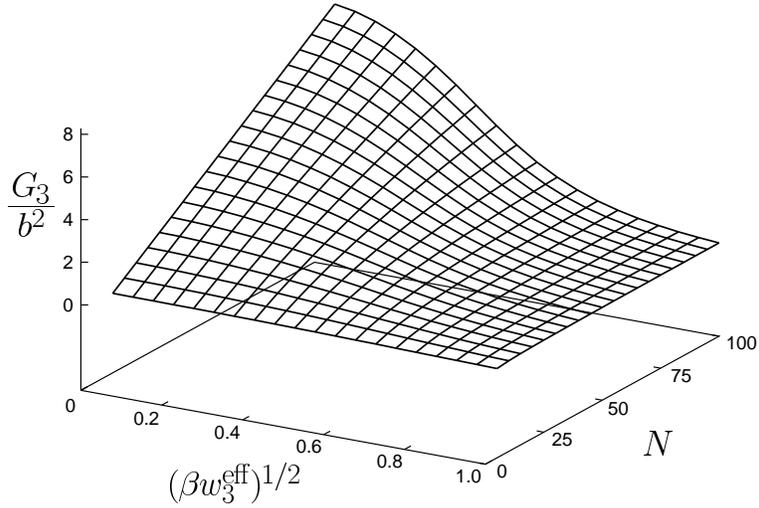,width=10cm}}} 
  \vspace*{10pt}
  \caption{\label{fig:adso1} Numerical result for the optimized 
    variational parameter $G_3\simeq \bar{R}_{\bot}^2$ as a function of
    interaction strength $(\beta w_3^{\text{eff}})^{1/2}$ and chain length
    $N$.  The localization transition can be identified: for small values of
    $w_3^{\text{eff}}$, $G_3$ grows linearly with $N$, whereas $G_3$ is
    independent of $N$ for large values of $w_3^{\text{eff}}$.}
\end{minipage}
\end{center}
\end{figure}

Thus in the limit of small effective interaction strength the chain has
Gaussian conformation (see figure \ref{fig:adso1}), whereas for high $\beta
w_3^{\text{eff}}$ the chain is localized at the surface, leading to a mean
polymer size which in lowest order shows the same characteristics as the
result of the scaling argument, equation (\ref{flat}).  From the conditions
for the limiting cases, a localization criterion 
\begin{equation} \label{loccrit}
  \beta\,w_{j\,\text{crit}}^{\text{eff}}\approx N^{-1/2}
\end{equation}
can be found, which means
$\beta\, w_{j\,\text{crit}}^{\text{eff}} \approx 0$ for long chains.
Therefore very long chains always are adsorbed, i.e.\ localized, at an
attractive surface.  This of course is a consequence of our assumption of a
penetrable surface, since in the opposite case of impenetrable surfaces
adsorption takes place only from a finite interaction strength \cite{Eisen},
i.e.\ $\beta\,w_{j\,\text{crit}}^{\text{eff}} > 0$.

For negative effective interaction strength $w_j^{\text{eff}}<0$, only the
case $\beta |w_j^{\text{eff}}| \ll N^{-1/2}$ is important for us, as we
are mainly interested in the adsorption behavior.  Expansion of
(\ref{Gallgemein}) in this case yields
\begin{equation}  \label{Rjminus}
  \bar{R}_j \simeq \sqrt{G_j} \simeq 
   R_0 \left\{1+c N^{1/2} \beta w_j^{\mathrm{eff}}\right\}. 
\end{equation}

\subsection{Effective Interaction Strength}

The full general form of the effective interaction strength is
\begin{eqnarray}
w_j^{\text{eff}}
&=& {G_j^{1/2}\over 2\pi(G_1G_2G_3)^{1/2}} 
    \int\dq x\, K[h(\ux)]\; w(\ux)  
    \left( 1 - {(B_j-h_j(\ux))^{2}\over G_j} \right) \,  
    \exp\left\{ - \sum_{i=1}^3 {(B_i-h_i(\ux))^2  \over 2G_i}\right\}
\end{eqnarray}
where the translational parameter {\bf B} has to be chosen such that
${\bf\nabla}_{\bf B}{\cal W} (\bf{B},\bf{G}) = {\bf 0}$.  In the following the
surface is assumed to be symmetrical with respect to the coordinates $x_1$ and
$x_2$.  Then we immediately have $B_1\stackrel{!}{=}0$ and
$B_2\stackrel{!}{=}0$ as a solution of the minimization equation (\ref{MinB}).
For simplicity, we additionally assume the surface heterogeneity to depend
only on one space direction $x_1$, which means $w(\ux)\equiv w(x_1)$ and
$h(\ux)\equiv h(x_1)$.  Hence $w_2^{\text{eff}}=0$ and the polymer extension
into the direction of $x_2$ equals that of a Gaussian chain.  In this case the
expressions for $w_1^{\text{eff}}$ and $w_3^{\text{eff}}$ reduce to
\begin{eqnarray}
w_1^{\text{eff}}  \label{w1eff}
&=& {1\over (2\pi G_3)^{1/2}} \int\!{\text d} x\, K[h(x)]\; 
    w(x) \left( 1 - {x^2\over G_1} \right)   
    \exp\left\{ - {x^2 \over 2G_1} - {(B_3-h(x))^2  \over 2G_3}\right\},\\
w_3^{\text{eff}}  \label{w3eff}
&=& {1\over (2\pi G_1)^{1/2}} \int\!{\text d} x\, K[h(x)]\; 
    w(x)\,\exp\left\{ - {x^2 \over 2G_1} - {(B_3-h(x))^2  \over 2G_3}\right\}.
\end{eqnarray}
Now a straightforward calculation for various types of surface heterogeneity
is possible.
\begin{description}
\item[(1)] For a flat surface with energetical heterogeneity, $h(x)=h_0$, the
  minimization condition (\ref{MinB}) results in $B_3\stackrel{!}{=}0$, so
  that the centre of mass of the chain is located on the surface.  Inserting
  $w(x)=\int_{-\infty}^\infty {\text d}q\, \exp\{iqx\}\,\tilde{w}(q)$ leads to
  \begin{eqnarray}
  w_1^{\text{eff}}
  &=& {G_1^{3/2}\over G_3^{1/2}} \int\limits_{-\infty}^\infty\!{\text d} q\, 
      q^2\, \tilde{w}(q)\, \exp\left\{ - {G_1 q^2 \over 2}\right\},\\
  w_3^{\text{eff}}
  &=&  \int\limits_{-\infty}^\infty\!{\text d} q\, \tilde{w}(q)\, 
       \exp\left\{ - {G_1 q^2 \over 2}\right\}.
  \end{eqnarray}
  As can be seen from the notation $\tilde{w}(q) = w_0\,\delta(q) +
  \tilde{w}^\ast(q)$, the effective interaction strength parallel to the
  surface is independent of the mean interaction strength $w_0$.
\end{description}
  \begin{itemize}
  \item[$\bullet$] For a periodic interaction strength $\tilde{w}(q) =
    w_0\,\delta(q) + (A_w/2) \{\delta(q-f)+\delta(q+f)\}$ with amplitude $A_w$
    and wave number $f$, we have
    \begin{eqnarray}
    w_1^{\text{eff}}
    &=& {G_1^{3/2}\over G_3^{1/2}} A_w f^2 e^{- G_1 f^2/2}\\
    w_3^{\text{eff}} \label{w3perint}
    &=&  w_0 + A_w e^{- G_1 f^2/2}
    \end{eqnarray}
    Thus $w_3^{\text{eff}}$ takes on its maximum $w_0 + A_w$, if the
    wavelength of the heterogeneity exceeds the polymer size parallel to the
    surface, $\lambda\simeq f^{-1} \gg \bar{R}_1$, because in this case the
    polymer chain, which is located at a maximum of $w(x)$, does not notice
    the existence of the minima of the interaction strength.  In the inverse
    case, $f^{-1} \ll \bar{R}_1$, the fluctuations of $w(x)$ cannot be
    resolved any more, $w_3^{\text{eff}}$ is minimal and equals the mean
    interaction strength.  $w_1^{\text{eff}}$ is always small (leading to a
    polymer size $\bar{R}_1\simeq R_0$ parallel to the surface), except for
    the case of a period of the fluctuation fitting the size of the chain,
    $f^{-1}\approx \bar{R}_1$, and large $A_w$.
  \item[$\bullet$] A randomly distributed interaction strength is best handled
    by identifying the amplitude in Fourier space $\tilde{w}^\ast(q)$ with the
    square root of the spectral density $S(q)$, so that $\tilde{w}(q) =
    w_0\,\delta(q) + c^{-1}\,\Delta_w\, \exp\{-q^2 \xi^2 \}$ for a Gaussian
    distribution with variance $\Delta_w^2$, correlation width $\xi$ and
    constant $c=(2\pi)^{1/4}$. Then the effective interaction strengths read
    \begin{eqnarray}
    w_1^{\text{eff}}
    &=& {G_1^{3/2}\over G_3^{1/2}}
        {c\,\Delta_w \over (G_1+\xi^2)^{3/2}},\\
    w_3^{\text{eff}} \label{w3ranint}
    &=&  w_0 + {c\,\Delta_w \over (G_1+\xi^2)^{1/2}}.
    \end{eqnarray}
    The magnitude of the heterogeneity is determined by both $\xi$ and
    $\Delta_w$: the smaller the correlation width and larger the variance, the
    stronger are the fluctuations, which leads to an increase of the effective
    interaction strength.  The limiting cases perpendicular to the surface are
    \begin{equation}
    w_3^{\text{eff}} \approx
    \left\{\begin{array}{l@{\quad{\mathrm{for}\ \ }}l}
      \ w_0 + c\, \Delta_w\, \xi^{-1} & \xi \gg \bar{R}_1\ , \\[7pt]  
      \ w_0 + c\, \Delta_w\, \bar{R}_1^{-1} & \xi \ll \bar{R}_1\ .  \\[3pt] 
    \end{array}\right. 
    \end{equation}
  \end{itemize}
\begin{description}
\item[(2)] In the case of a heterogeneous surface profile (whereas the
  interaction strength $w(x)=w_0$ is assumed constant) the disorder has to be
  weak in order to make the $x$ integration feasible. Therefore we only
  investigate the case $|h(x)|\ll 1$ and $|\nabla h(x)|\ll 1$, where $\langle
  h(x)\rangle=0$, and restrict the calculation to first order in the
  fluctuation of $h(x)$.  With $h(x)=\int_0^\infty {\text d}q\,
  \cos(qx)\,\tilde{h}(q)$, the minimization (\ref{MinB}) yields $B_3\approx
  \int_0^\infty {\text d}q\, \tilde{h}(q)\, \exp\{-G_1q^2/2\}$.  This means
  that the centre of mass of the chain to some extent follows the surface
  profile.  For an example see figure \ref{fig:Bzmin1}, where $\textbf{B}$ is
  sketched for a periodic surface profile.
  \begin{figure}
    \begin{center}
    \begin{minipage}{12cm}
    \centerline{{\epsfig{file=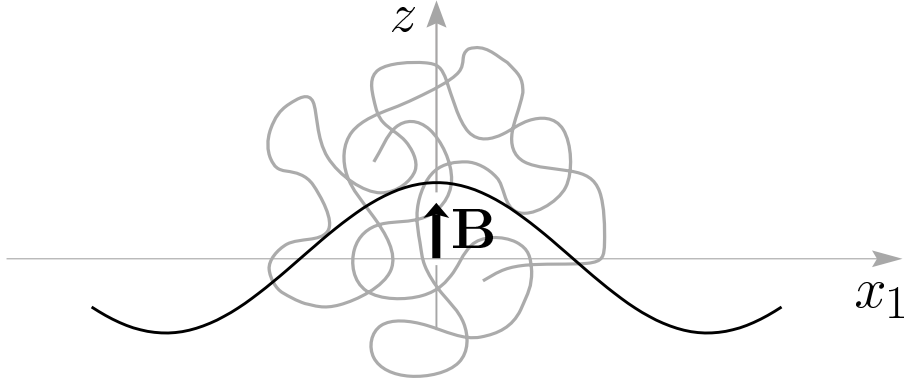,width=12cm}}} 
    \vspace*{10pt}
    \caption{\label{fig:Bzmin1} Sketch of the optimized translational shift
      ${\textbf B}$ of the centre of mass of the chain for a sinusoidal
      surface profile.}
    \end{minipage}
    \end{center}
  \end{figure}
  Now the deflection factor can be approximated by
  \begin{equation}
    K[h(x)]=(1+|\nabla h(x)|^2)^{1/2} \approx 1 + {1\over2} 
    \int\limits_0^\infty\!{\text d}q \int\limits_0^\infty\!{\text d}q^\prime\,
    \tilde{h}(q)\,\tilde{h}(q^\prime)\, qq^\prime\,\sin(qx)\,\sin(q^\prime x).
  \end{equation}
  If additionally the part of the exponent in (\ref{w1eff}) and (\ref{w3eff})
  which depends on $h(x)$ is expanded, we obtain in lowest order of
  $\tilde{h}(q)$
  \begin{eqnarray}
  w_1^{\text{eff}}
  &\approx& w_0\, \left({G_1\over G_3}\right)^{3/2} 
      \int\limits_0^\infty\!{\text d} q 
      \int\limits_0^\infty\!{\text d} q^\prime\,
      \tilde{h}(q)\,\tilde{h}(q^\prime)\,
      \exp\left\{ - {G_1 (q^2 + {q^\prime}^2) \over 2}\right\} \times
      \nonumber\\
  && \qquad\times\, \left[ q^2 + {(q+q^\prime)^2\over2}\left( 
            G_3 q q^\prime \sinh(G_1qq^\prime) - \cosh(G_1qq^\prime) 
       \right)\right]\\
  w_3^{\text{eff}}
  &\approx&  w_0 \left\{ 1 + {1\over2G_3} \int\limits_0^\infty\!{\text d} q 
      \int\limits_0^\infty\!{\text d} q^\prime\,
      \tilde{h}(q)\,\tilde{h}(q^\prime)\,  
      \exp\left\{ - {G_1 (q^2 + {q^\prime}^2) \over 2}\right\} \times
      \right. \nonumber\\
  && \hspace*{3cm}\times\,  \Bigl[ 3 + 
      G_3qq^\prime \sinh(G_1qq^\prime) - \cosh(G_1qq^\prime) \Bigr] \Biggr\}
  \end{eqnarray}
\end{description}
  \begin{itemize}
  \item[$\bullet$] A periodic surface geometry $\tilde{h}(q)=A_h\delta(q-f)$
    leads to
    \begin{eqnarray} 
      w_1^{\text{eff}} \label{w1pergeo}
      &\approx& w_0\, \left({G_1\over G_3}\right)^{3/2} {A_h^2f^2\over2} 
        \left\{ G_3f^2 \left(1-e^{-2G_1f^2}\right) 
        -  \left(1-e^{-G_1f^2}\right)^2 \right\}\\
      w_3^{\text{eff}} \label{w3pergeo}
      &\approx& w_0 \left\{ 1 
        + {A_h^2f^2\over 4} \left(1-e^{-2G_1f^2}\right) 
        - {3A_h^2\over 4G_3} \left(1-e^{-G_1f^2}\right)^2 \right\}
    \end{eqnarray}
    For a flat surface, the polymer extension $\bar{R}_3$ parallel to the
    mean surface normal (i.e.\ in $z$ direction) is identical to the size
    perpendicular to the surface.  This is different for a rough surface
    profile, which now turns out to be important for the interpretation of
    (\ref{w1pergeo}) and (\ref{w3pergeo}).  If for example $G_3$ is small
    compared with the squared wavelength $\lambda^2\simeq f^2$, then the
    polymer sticks to the surface, following the deflections.  Therefore the
    result for $\bar{R}_3\simeq\sqrt{G_3}$ exceeds the polymer size
    perpendicular to the surface by the amount of the deflection, the
    effective interaction strength $w_3^{\text{eff}}$ accordingly is smaller
    than $w_0$.  Adsorption enhancement therefore only can be obtained in the
    opposite case $\bar{R}_3\gg f^{-1}$, where the effective interaction
    strength parallel to the surface normal takes on its maximum value
    $w_{3\ \text{max}}^{\text{eff}} = w_0 \{ 1 + A_h^2f^2/4\}$.
    
    Here we have used $G_1\leq G_3$, which results from the fact that
    $w_1^{\text{eff}}$ always is small according to the condition $A_hf\ll 1$.
    As can be seen from (\ref{w1pergeo}), in the case $G_3\ll f^{-1}$
    discussed above $w_1^{\text{eff}}$ is negative, so that the mean polymer
    extension in $x$ direction exceeds the size of a corresponding Gaussian
    chain, see (\ref{Rjminus}).
  \item[$\bullet$] Similar to the case of a randomly distributed interaction
    strength, the amplitude in Fourier space $\tilde{w}^\ast(q)$ of a randomly
    distributed surface profile is identified with the square root of the
    spectral density $S(q)$.  This means $\tilde{h}(q) = c^{-1}\, \Delta_h\,
    \exp\{-q^2 \xi^2/2 \}$ for a Gaussian distribution with mean $0$, variance
    $\Delta_h^2$ and correlation width $\xi$.  In order to satisfy the
    requirement of weak disorder, we have to assume $\Delta_h^2\ll 1$ and
    $\xi\gg 0$.  Then the result for the effective interaction strength
    in $z$ direction is
    \begin{eqnarray}
     w_3^{\text{eff}} 
     &=& w_0\left\{ 1 + \sqrt{{\pi\over 2}} {\Delta_h^2\over 4}
     \left[ {G_1\over\xi^3} (2G_1+\xi^2)^{-3/2} 
     + {3\over G_3}\left( (G_1+\xi^2)^{-1} 
       - {1\over\xi} (2G_1+\xi^2)^{-1/2}  \right) \right] \right\}.
    \end{eqnarray}
    For a very large correlation width, which in the limit $\xi\to\infty$
    corresponds to a flat surface, we again have the effect of a reduction of
    the effective interaction strength compared with the flat surface,
    $w_3^{\text{eff}}<w_0$.  Therefore the result which is relevant for
    adsorption enhancement here is obtained in the case $\Delta_h\leq\xi^2\ll
    G_3\leq G_1$, where the effective interaction strength has its maximum
    value
    \begin{equation} \label{w3rangeo}
    w_{3\ \text{max}}^{\text{eff}} \approx
    \ w_0 \left\{ 1 + {\textstyle c\,\Delta_h^2 \over 
       \textstyle\xi^3 \sqrt{G_1} } \right\}.     
    \end{equation}
    An analytic expression for $w_1^{\text{eff}}$ is not available, but the
    main features of the result can be estimated to strongly resemble those of
    $w_1^{\text{eff}}$ for a periodic surface profile discussed above.
  \end{itemize}

\section{Conclusions} \label{conc}

The variational calculation presented here is valid for weak spatial disorder
only (therefore it does not reproduce the scaling behavior for fractal
surfaces).  Nevertheless the mechanism of adsorption enhancement is well
reproduced, we find agreement of the results in all special cases which were
already investigated in the literature.

A special feature of the variational method employed here is the possibility
of quantifying the localization transition, i.e., the transition from a
slightly deformed Gaussian coil to a localized conformation, where the polymer
size perpendicular to the surface no longer depends on the chain length.
According to the result (\ref{loccrit}) the localization can be obtained by
increase either of the effective interaction strength or of the chain length.
This helps to compare the strength of adsorption enhancement for the two sorts
of disorder considered here: as can be seen from the maximum values of
(\ref{w3perint}) and (\ref{w3pergeo}) or from a comparison of (\ref{w3ranint})
and (\ref{w3rangeo}), the localization transition is only slightly affected by
a rough surface profile, whereas energetical heterogeneity can induce the
transition even at vanishing mean interaction strength.  Therefore we conclude
that the disorder-induced enhancement of polymer adsorption is much more
significant for a heterogeneous interaction strength than for spatial
roughness.

Our findings concerning the localization behavior are affected qualitatively
by the assumption of surface penetrability: for infinitely long chains at a
flat and homogeneous impenetrable surface, the localization transition in
contrast to (\ref{loccrit}) only occurs for some nonzero value of the
attractive potential \cite{Eisen}.  Nevertheless we expect our main statement
on the significance of adsorption enhancement to hold also for impenetrable
surfaces, since the comparison of transparent and opaque surfaces in simple
soluble cases by Hone et al.\ \cite{HJP} shows that they should not be
affected differently by weak surface heterogeneities.

The polymer size parallel to the surface does not directly depend on the mean
interaction strength, but only through the extension perpendicular to the
surface. Thus, because it is less affected by heterogeneity, the former always
exceeds the latter, except for one special case: for a flat, neutral surface
with a periodic interaction strength, the polymer size parallel to the surface
is smaller than perpendicular to it, if the period fits the polymer size such
that it is concentrated to a maximum of $w(x)$ and even restricted by the
neighboring repulsive regions.

\section{Acknowledgments}

We would like to thank G.\ Heinrich for stimulating discussions.
Financial support by the Deutsche Kautschuk Gesellschaft (DKG) is gratefully
acknowledged.


\vfill\break

\end{document}